\documentclass[journal,10pt,twocolumn,twoside,letterpaper]{IEEEtran}

\usepackage[latin1]{inputenc}

\usepackage{mdwlist}
\usepackage[cmex10]{amsmath} %
\usepackage{amsthm,amssymb}

\usepackage[pdftex]{graphicx} 
\usepackage[colorlinks=false]{hyperref}
\usepackage{sidecap}

\usepackage{cite}
\def\ArticleTitle{Hierarchical Recursive Running Median}

\def\RightPageHeading{A.~Alekseychuk: \ArticleTitle}

\hypersetup
{
  unicode,
  bookmarks=true,
  bookmarksopen=true,
  pdfstartview=Fit,
  pdfpagelayout=SinglePage,
  pdftitle=\ArticleTitle, 
  pdfsubject={fast algorithm for calculation of running median}, 
  pdfkeywords={median filter, running median,
    recursive algorithms, energy efficient algorithms, fast
    algorithms, computational complexity}, 
  pdfauthor={Alexander Alekseychuk (alexander.alekseychuk@tu-berlin.de, alex.alekseychuk@gmail.com)} 
}

\begin{document}

\author
{
  Alexander~Alekseychuk%
\thanks{
  A.~Alekseychuk is with the 
  Computer Vision and Remote Sensing Group at Technische Universität Berlin, Germany, 
  phone:~+49~30~31473109, 
  email:~alexander.alekseychuk@tu-berlin.de or alex.alekseychuk@gmail.com}%
\thanks{
  This work was supported by the German Federal State of Berlin in the
  framework of the ``Zukunftsfonds Berlin'' and the Technology Foundation
  Innovation center Berlin (TSB) within the project ``Virtual Specimen
  Scout''. It was co-financed by the European Union within the European
  Regional Development Fund (ERDF).}
}

\title{\ArticleTitle}

\maketitle

\begin{abstract}

To date, the histogram-based running median filter of Perreault and
H\'{e}bert is considered the fastest for 8-bit images, being roughly
$O(1)$ in average case. We present here another approximately constant
time algorithm which further improves the aforementioned one and
exhibits lower associated constant, being at the time of writing
the lowest theoretical complexity algorithm for calculation of 2D and
higher dimensional median filters.  The algorithm scales naturally to
higher precision (e.g. 16-bit) integer data without any modifications.
Its adaptive version offers additional speed-up for images showing
compact modes in gray-value distribution.  
The experimental comparison to the previous
constant-time algorithm defines the application domain of this new
development, besides theoretical interest, as high bit depth data
and/or hardware without SIMD extensions.  The C/C++ implementation of
the algorithm is available under GPL for research purposes.

\end{abstract}


\ifCLASSOPTIONdraftcls
\begin{center} \bfseries EDICS category: TEC-PRC \end{center}
\else
  \ifCLASSOPTIONpeerreview
  \begin{center} \bfseries EDICS category: TEC-PRC \end{center}
  \fi
\fi

\markboth{}%
{\RightPageHeading}

\section{Introduction}

\IEEEPARstart{M}{edian} filter was long known in image processing for
its high computational cost.  Reduction of computational complexity of
median filter was tackled many times in course of years.  In this process the
computational complexity of $O(n^2)$ of straight forward
implementation was gradually reduced to $O(n)$ in~\cite{Huang1979},
$O(\log^2{n})$ in~\cite{Gil1993} and $O(\log{n})$ in~\cite{Weiss2006},
where $n$ is the filter size, i.e. the size of one side of square 2D
filter window.  These were still relatively high values in
comparison to efficient implementations of some separable linear
filters, where $O(1)$ complexity with respect to filter size $n$ can be
reached.  Finally in 2007 a first algorithm for computation of 2D
median exhibiting roughly $O(1)$ average case complexity was
proposed~\cite{Perreault2007}.  Although this algorithm is able to
compute running median in constant average time per pixel of output
image, the associated constant is still significant and thus it relies
on SIMD (single instruction, multiple data) extensions of modern CPUs
along with some data-dependent heuristics to lower this value.

We present here another constant time median filtering algorithm 
combining ideas of search trees with histogram-based approaches
of~\cite{Weiss2006} and~\cite{Perreault2007}.  This algorithm exhibits
lower associated constant in terms of necessary number of operations than the
aforementioned ones, being, to our best knowledge at the time of
writing, the lowest computational complexity algorithm for calculation
of 2D and higher dimensional median filters.  Two versions of the
algorithm are presented.  The first one is appropriate if no a priori
knowledge about data is available.  The second version of the
algorithm
further improves efficiency
for images showing compact modes in gray-value distribution.
This is the case, for example, for images with prevailing gray value
ranges such as low-key or high-key images, images taken under
insufficient illumination, X-ray images of some types, etc.

The developed algorithm 
makes approximately two times less operations on 8-bit data than the
best previous algorithm.  This is a significant improvement since the
competitive algorithm has already an $O(1)$ average case complexity.
However, despite the algorithm
makes in fact lower number of operations, it cannot benefit much of
SIMD extensions of current CPUs.  Thus, despite higher efficiency
, we were not able to show 
practically relevant advantage on 8-bit data over the older
algorithm~\cite{Perreault2007} of Perreault and H\'{e}bert if executed
on a current main-stream CPU.  Nevertheless, the algorithm can be
interesting not only from the theoretical point of view, but seems to be the
best choice on platforms lacking hardware SIMD extensions,
e.g. embedded systems, mobile devices, etc. as well as for higher
precision data like, for example, HDR and X-ray imaging.  The shortage
of SIMD utilization also does not prevent usual higher-level
parallelizations of the algorithm.

In the following section we give short overview of important known
approaches to calculation of median filter and analyze their
advantages and disadvantages.  In section \ref{AlgDescriptionSection}
we describe in details the proposed algorithm.  Finally, results of
experimental evaluation are presented.  The developed algorithm
produces the same filtering results as any other median filter.
Therefore we deliberately show no filtering results, but concentrate
on comparison of both versions of the developed algorithm to the
constant time algorithm~\cite{Perreault2007} of Perreault and
H\'{e}bert in terms of performed number of operations and computing
time.

\section{Overview and analysis of known approaches}

Median element $m$ of a finite ordered set $S$ can be defined as the
smallest element such that a half of elements in $S$ are less than or
equal to $m$.
In image processing the
set $S$ is created by specifying a rectangular window, or in general
case an arbitrary shaped mask, centering the window at a particular
image point and enumerating all pixel values located inside the
window.  Then the median of these values is found and used as 
filter output at the given point.  In 2D median filter, also known
as running or moving median, this procedure is repeated for each image
point.

The straight-forward implementation of median filter follows
the procedure described above, finding the median by sorting all
values inside the current $n \times n$ filter window and
taking the value located at $n^2/2$ position.
Application of {\em Quicksort}~\cite{Hoare1961} as the sorting
algorithm results in computational complexity of $O(n^2\log{n})$
operations per pixel of result image.  As a side-effect, any other
order statistic can be found after that in an $O(1)$ time since
pixel values inside the window are sorted already.

Instead of using full sort, an algorithm known as {\em
  Quickselect}~\cite{Floyd1975} can be used.  It does not sort
the complete buffer but only places one $k$-th smallest, in case of median
the $(n^2/2)$-th smallest, element at place it would occupy after
sorting.  This is just sufficient if only one order statistic is of
interest, e.g. only median has to be calculated.  Its computational
complexity is $O(n^2)$ with everything else being the same as for the
full sort method.

These approaches share one major drawback: results of median
calculation in one image point can not be utilized by {\em Quicksort}
and {\em Quickselect} algorithms for finding the median at the next
window position.  Instead, all steps, i.e. selection of window points,
copying them into temporal buffer and, finally, sorting them, have to be
repeated from scratch for each pixel of the resulting image.  Whereas
sorting is by far the most time consuming operation in this chain.

\begin{figure}
\centering \includegraphics[width=0.7\linewidth]{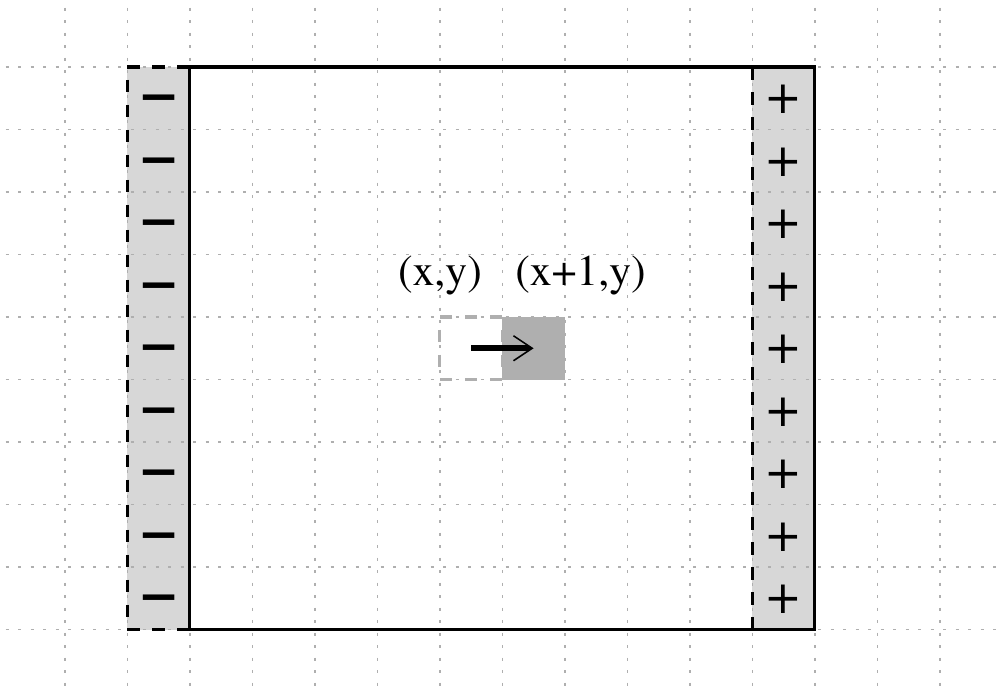}
\caption{Basic principle of a rectangular-shaped 2D filter
  implementing 1D recursion}
\label{1d-recursion-principle}
\vspace{-10pt}
\end{figure}

Search trees can be used in order to address this problem, i.e. to
reuse sorting result at previous window position.  
Note that rectangular windows at positions $(x,y)$ and $(x+1,y)$
overlap to a great extend~(Fig.~\ref{1d-recursion-principle}),
i.e. share most of their pixels.  Having a search tree filled with
pixels values from window at $(x,y)$, the move to position $(x+1,y)$
is performed by removing $n$ pixel values with coordinates $(x-n/2-1,
y_i)$ and adding $n$ pixel values with coordinates $(x+n/2, y_i)$,
where $y_i \in [y-n/2, y+n/2]$.  In case if self-balancing binary
search trees are used, an insertion or removal can be done in average
$O(\log{n})$ and median extraction in constant $O(1)$ time.  The
overall average computational complexity of the above algorithm is
therefore $O(n\log{n})$ per pixel of the resulting image.

Gil and Werman took in~\cite{Gil1993} a similar
approach processing the image in blocks of $(2n-1)^2$ pixels.  They
store all $(2n-1)^2$ pixels in a special search-tree based data
structure that they call {\em Implicit Offline Balanced Binary Search
  Tree} (IOBBST).  In fact, they use two nested IOBBST, i.e. {\em each}
node of the main IOBBST contains a secondary IOBBST again.  They do
not rebuild this data structure as the running window moves inside the
block but mark respective pixels as {\em active} or {\em inactive}.
The algorithm is able to find $n^2$ medians in $O(n^2\log^2{n})$ time,
further reducing the computational complexity of median calculation down
to $O(\log^2{n})$ per pixel.

We are not aware of any implementation of Gil and Werman
algorithm as well as no experimental evaluation was given in their
article.  
However, the complexity of the algorithm flow suggests that the 
theoretical $O(\log^2{n})$ complexity translates to the real-world running time 
with a high (constant) factor. 

The positive property of sorting-based approaches is that no
assumptions about image data have to be made.  A completely different
approach is based on a histogram of pixel values.
The median can be found by successive summing
of histogram bins in increasing order until the sum reaches $n^2/2$.  
No difference between pixels
falling into the same gray value range is made in this case.  This can be 
beneficial for large $n$ and low number of histogram bins.

The Tibshirani's {\em Binmedian} algorithm~\cite{Tibshirani2009}
belongs to such histogram-based algorithms, although it is not
specifically intended for 2D image processing.  It relies additionally
on the Chebyshev's inequality stating that the
difference between the median $m$ and the mean $\mu$ is always at most one
standard deviation: $ |\mu-m| \le \sigma$.
The basic idea is to build a low resolution histogram of values in
$[\mu-\sigma, \mu+\sigma]$ interval, i.e. map data values to bins
inside this interval, find the bin which contains the median by
successive summing of bins and then recur with finer histogram
on values inside this bin until the required precision is reached.  
Relaxing precision requirements by not performing recursion on the
median bin gives us the approximate version of {\em Binmedian}
algorithm: {\em Binapprox}.  It is accurate up to $1/B^{th}$ part of
standard deviation, where $B$ is number of histogram bins.

The average case complexity of {\em Binmedian} algorithm is $O(n^2)$
and the computational complexity of {\em Binapprox} was estimated as
$O(n^2)$ for the worst case.  If applied to each image point
independently they offer no significant advantage over the classic {\em
  Quickselect}.  However, {\em Binmedian} and {\em Binapprox}
algorithms can be used in update mode, reusing the
histogram calculated at previous window position and only seldom resorting
to full recomputation.  An order of magnitude speedup was reported for
this case~\cite{Tibshirani2009}.

All algorithms described above can work on data with arbitrary
number of quantization steps or even floating point data.  Further
performance improvement can be made if fixed and a priori known
gray value resolution is assumed.  This is the case for
most digital images and is especially advantageous for broadly used
256-values per channel images.  

This approach is taken in the Huang et al. algorithm~\cite{Huang1979}
where a 256-bin histogram is used for counting of gray values in a 2D
moving window.  The median calculation is performed in the same way as
above by successive summing of individual histogram bins until the
median condition is reached.  This algorithm utilizes the overlap
between two 2D windows at neighboring positions and updates the window
histogram recursively.  At each subsequent window position only values
of the new pixel column are added to and values of the obsolete
column are subtracted from the histogram calculated at previous
position.  This results in $O(n)$ computational complexity.
Nevertheless the recursion is only one-dimensional and there is still
room for improvements.

These have been done by Weiss in~\cite{Weiss2006} who has improved
Huang's algorithm by using the distributive property of histograms.
Weiss's approach is to process several ($2n^{1/2}$, $4n^{2/3}$ up to
$O(n)$) image rows at the same time and to maintain a set of partial
histograms instead of using only one.  These partial histograms
reflect smaller image areas, than one single histogram would do, and
thus, can be updated more efficiently.  They are implicitly combined
to a single histogram for final histogram-based median calculation.
The same set of partial histograms is used for all rows which are
processed simultaneously.  Thus much of redundancy of Huang's
straightforward algorithm is avoided and average cost for histogram
maintenance is lowered.  Overall computational complexity becomes
$O(\log{n})$.

Weiss's algorithm can be also adapted to higher than 8-bit gray value
resolutions via a technique called {\em ordinal transform}.  It
consists in sorting of {\em all} pixel values appearing in a given
image and replacing them with their order-value.  This allows more
efficient histogram representation and associated arithmetic, saving
space required for storage of histograms, but giving the algorithm
$O(\log^2{n})$ complexity.

Perreault and H\'{e}bert~\cite{Perreault2007} have chosen another way
for improvement of Huang's approach.
They use separate histograms for each column of the moving window
(which moves in horizontal direction).  The column histograms are
cached and can be efficiently updated in a recursive way.  The window
histogram is then created by summing of respective bins of column
histograms.  Since this is a linear operation, the window histogram
can be recursively computed too.  This requires subtraction of the
column histogram which went out of scope of the running window and
addition of the histogram for the column which was newly added to the
window (Fig.~\ref{1d-recursion-principle}).  Both operations are
constant time, i.e. the number of operations is independent of window
size $n$.  Albeit this number is significant, e.g. for 8-bit data it
is 256 additions and 256 subtractions per resulting pixel.  The
calculation of median by summing of histogram bins also requires in
average a constant time (127 additions and 128 comparisons in case of
8-bit data).  Thus, neglecting initialization of column histograms for
the first row and initial creation of the window histogram at the
beginning of each row, which are $O(n)$ operations, the overall
algorithm also becomes constant time.  The high associated constant
ought to be compensated by intensive usage of SIMD extensions (single
instruction, multiple data) of modern vectorized CPUs.  They also
propose several heuristics for reducing of the high associated
constant.  The most powerful among them is to delay summation of
column bins for the window histogram, till the respective bin is
actually required for median calculation, i.e. to perform summation
on-demand.

Another important refinement, applied by Perreault and H\'{e}bert (but
first appeared in~\cite{Alparone1994} by Alparone et al.), consists in
usage of two-tier window histogram.  The higher tier is called coarse
level and consists of reduced number of bins (16 in this case), while
the lower tier contains the usual full resolution histogram (256
bins).  Such scheme allows to find the median faster by first scanning
the low-resolution part and then continuing only in a limited range of
the full-resolution histogram.  Tibshirani's successive binning
algorithm~\cite{Tibshirani2009} uses basically the same idea.  It
allows to reduce the average number of operations for finding median
from 127 additions and 128 comparisons in case of 8-bit single-level
histogram to roughly 16 additions and 16 comparisons using two-level
histogram.

Note, the use of two-level histograms in Perreault and H\'{e}bert
algorithm results in more important effect than just faster extraction
of median.  This procedure allows to reduce the number of potentially
computationally intensive updates of bins in the window histogram
because they are updated on-demand.  Perreault and H\'{e}bert have
analyzed computational complexity of their algorithm only for the
version with full unconditional histogram update whereas the on-demand
version
is of practical interest.  It was not done since it is difficult to
track analytically.  Our experimental results show that updates of
window histograms are indeed the most time consuming part in Perreault
and H\'{e}bert algorithm.  This is an important observation and we
fully develop an idea which addresses this problem in our algorithm.

\section{Hierarchical recursive running median}
\label{AlgDescriptionSection}

The main idea of the proposed algorithm is to optimize the calculation
of various order statistics, and median among them, using a
special data structure which we call {\em interval-occurrences tree} (IOT).

\begin{figure} %
\centering \includegraphics[width=1.0\linewidth]{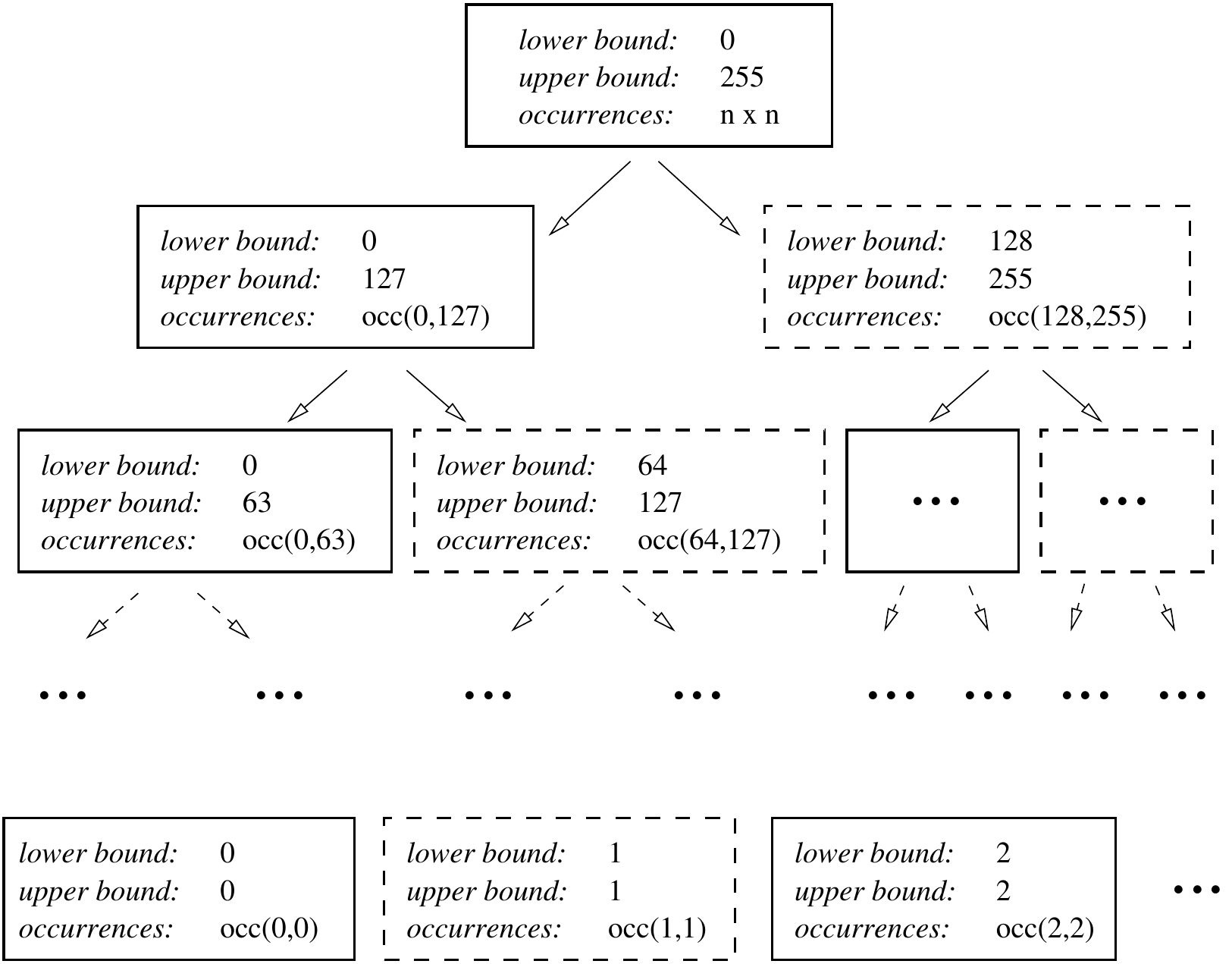}
\caption{{\em Interval-occurrences tree} data structure. Each node has
  associated lower and upper bounds ($b_{lo}$ and $b_{up}$), which
  define an interval of pixel values, and stores the number of
  occurrences $occ(b_{lo},b_{up})$ of pixels with gray values
  inside this interval.  Intervals and occurrences of child
  nodes sum to the interval and occurrences of their parent. Leaf
  nodes correspond to smallest possible (or smallest required) value
  intervals, e.g. 1 gray value in case of integer data.}
\label{explicit-iot-example}
\vspace{-10pt}
\end{figure}

An interval of pixel values, defined by the lower and the upper bound,
is associated with each node of the IOT.  Each IOT node stores the number of
occurrences of pixels which gray values belong to these intervals.  An
IOT is build in the following way (Fig.~\ref{explicit-iot-example}):
\begin{itemize*}
\item The top-most node has minimal possible pixel value as lower
  bound and maximal possible pixel value as higher bound.
  Number of occurrences is equal to the overall number of pixels in the
  region which is described by the IOT, i.e. all pixels are included.
\item Each node has exactly two children.  They subdivide
  the interval of the parent node in two sub-intervals and store
  corresponding number of value-occurrences in each sub-interval.
\item Other nodes are defined recursively until the value interval
  associated with the node vanishes.  For example, in case of integer
  data leaf nodes correspond to intervals of 1 gray value.
  Alternatively, the tree building can be stopped as soon as a
  required gray value precision is reached and thus floating point
  values can be processed too.
\end{itemize*}

Obviously, it is necessary to define {\em how} 
child-nodes divide
the interval of their parent.  The simplest rule would be to 
divide it in two equal-length parts.  We will call such
IOT a {\em uniform interval-occurrences tree}.  
Uniform IOT of height $d+1$ is necessary to represent integer data
with effective bit-depth $d$ up to the precision of one gray value.  
We will address later how such an IOT
can be created and stored in memory in an efficient way as well as how
to fill and update it with pixel values.

Given a uniform IOT representation, any order statistic $k$ can be found
by visiting at most $d$ nodes, requiring $d$
comparisons and in average $d/2$ additions.

\begin{IEEEproof}
Consider the following algorithm for finding the smallest lower value interval with
number of pixel occurrences equal or greater than $K=k n^2$, where
$k$ is the requested percentile:

{\em Get\_order\_statistic} procedure: \label{extract-median-algo}
\begin{enumerate*}
\item Make topmost node the current one, set {\em
  left\_occurrences\_accumulator} to zero.
\item \label{extract-median-algo-loop} If number of occurrences in the
  left node (lower sub-interval) plus {\em
    left\_occurrences\_accumulator} is greater 
  than the requested value $K$, then descend to the left sub-tree
  (Fig.~\ref{explicit-iot-example}), i.e. the sub-tree which further
  subdivides the left sub-interval.  For this: make the left child
  node the current one.  Otherwise: add number of occurrences in the
  left node (lower sub-interval) to the {\em
    left\_occurrences\_accumulator} and descend to the right sub-tree,
  making the right child node the current one.
\item If value interval corresponding to the current node is greater
  than the required precision, then repeat step
  \ref{extract-median-algo-loop}.  Otherwise: return the 
  lower interval bound corresponding to the current node.
\end{enumerate*}

The procedure above visits at most 8 nodes and makes at most 8
descends to sub-trees for 8-bit data and at most 16 nodes and descends
for 16-bit data because these are the heights of the respective trees
(excluding the topmost node).  The returned value is equal, up to the
required precision, to the value of $K$-th element of the list of
individual pixel values sorted in ascending order and therefore it is
the $k$-th order statistic by definition.
\end{IEEEproof}

The median filtering algorithm uses one IOT for the running window and
separate IOTs of the same structure for each image column of size $n$
(assuming the running window moves in horizontal-first way).  Same
structure of two IOTs means that: (a) they are constructed for the same
initial value interval, e.g. 0-255 gray values; (b) use the same rule
for subdivision of the parent intervals into child intervals and
(c) are built up to the same gray value resolution, e.g. 1 gray value.
As the running window advances in horizontal direction its IOT is
updated using column IOTs in a recursive way.  
That is: the number of value occurrences in a particular node of window IOT 
is decremented by the number of occurrences in the corresponding node
of the column IOT which just went out of scope of the window and
incremented by the value from the corresponding node in the column IOT
which was just included in the window.

Node updates described above can be done for the entire window IOT.
The computational complexity is proportional to the IOT size but
independent of window size $n$, thus, it is an $O(1)$ operation with
respect to $n$.

Instead of updating the entire window IOT, an on-demand update can be
performed similarly as in~\cite{Perreault2007} by
updating only nodes actually required for median calculation at a
given position.  
At most $d$ updates will be necessary since any order statistic can be
found by visiting of $d$ nodes.  
Note that nodes on higher levels store occurrences corresponding to
larger gray value intervals.
The {\em get\_order\_statistic} procedure will with high
probability visit the same nodes at the next window position.  Such
updates will require just one addition and one subtraction per
visited node, i.e. are independent of $n$.  
We will call such updates {\em elementary}.
Updates of nodes which were not visited since some window advances
will require, however, execution of all delayed subtractions and
additions, up to recalculation from scratch, requiring at most $n$
additions of occurrence counters in corresponding nodes of column
IOTs.

As the running window advances to next image line, the column IOTs
have to be updated too.  This is also performed in a recursive way:
new pixel values are added and obsolete values are subtracted from
corresponding column IOTs, resulting in one added and one removed
pixel per each column IOT.  These are constant time operations too (exact
description in next subsections).

One can see that the algorithm implements 2D recursion.  The
computational complexity of the version doing unconditional updates of
the entire window IOT is independent of the window size $n$ for average as
well as for the worst case.  On-demand updates of the window IOT allow
to further reduce the algorithm complexity for the average case.

Indeed, due to the hierarchical IOT design most updates are elementary
(in our experiments with average data and $n \simeq
20..50$ about 95\% of updates are elementary).
Larger $n$ usually decreases the number of ``costly'' non-elementary
updates because of smoother result.  
Upper complexity bound for a non-elementary update grows, however, with
$n$.  

Empirical data show that these two effects successfully compensate
each other.  Thus, the complexity of the algorithm doing on-demand
update of the window IOT remains in average only {\em approximately}
constant.  The associated factor is, however, significantly lower than
for the version which does unconditional updates of the entire window
IOT and which constitute the upper complexity bound, strictly $O(1)$
but with a higher associated constant.

The proposed algorithm can be easily extended to N-dimensional case.
Whereas the same is true for~\cite{Perreault2007}, the advantage in
computational complexity of the developed algorithm will be higher and
grow proportionally to data dimensionality.

\subsection{Implicit interval-occurrences tree and memory requirements}

All IOTs in the above algorithm have the same structure.  Therefore,
lower and upper bounds of nodes' sub-intervals can be stored only once
in a single look-up table and need not to be repeated for each IOT.
\footnote{ Implementation note:
  On most hardwares it is even more efficient to have no look-up table
  for uniform IOT interval bounds, but calculate them on-fly.  This
  will require only one binary shift and in average $1/2$ integer
  addition per descend-step (assuming the IOT is built for gray value range
  ending at power of 2 boundary, e.g. 0-255 or 0-$(2^{16}-1)$).
  }

A further significant improvement can be made by closer look at the {\em
  get\_order\_statistic} procedure.  One can see that only the number
of pixel occurrences stored in left nodes are used for order statistic
calculation.  This suggests that right nodes (shown on
Fig.~\ref{explicit-iot-example} with dashed line) need not to be
stored in the IOT at all.  Note that child nodes describe two value
sub-intervals which build together the parent's interval.  In
case the occurrences in the right sub-interval will be required at
some later moment of time, they can be easily calculated by subtraction
of the left sub-interval occurrences from the parent interval
occurrences.  Leaving out right nodes not only saves space, but, as it
is shown in the next section, also saves computations during insertion
and removal of pixel values in an IOT.

We call an IOT without right nodes {\em implicit interval-occurrences
  tree}.  The number of nodes in an implicit IOT is calculated as sum
of geometric series with common ratio of 2 and equals the size of a
plain histogram for the same gray value range: $N = 1 +
\sum_{i=1}^d{2^i/2} = 1 + \sum_{i=0}^{(d-1)}{2^i} = 2^d$, where $1$ in
the first part of this formula is for the top-most node, $i$ iterates
from $1$ to $d$ because the tree has $d$ layers without the top-most
node and the expression behind the sum sign is divided by 2 because
only left nodes are stored explicitly.

\begin{figure} 
\centering \includegraphics[width=0.95\linewidth]{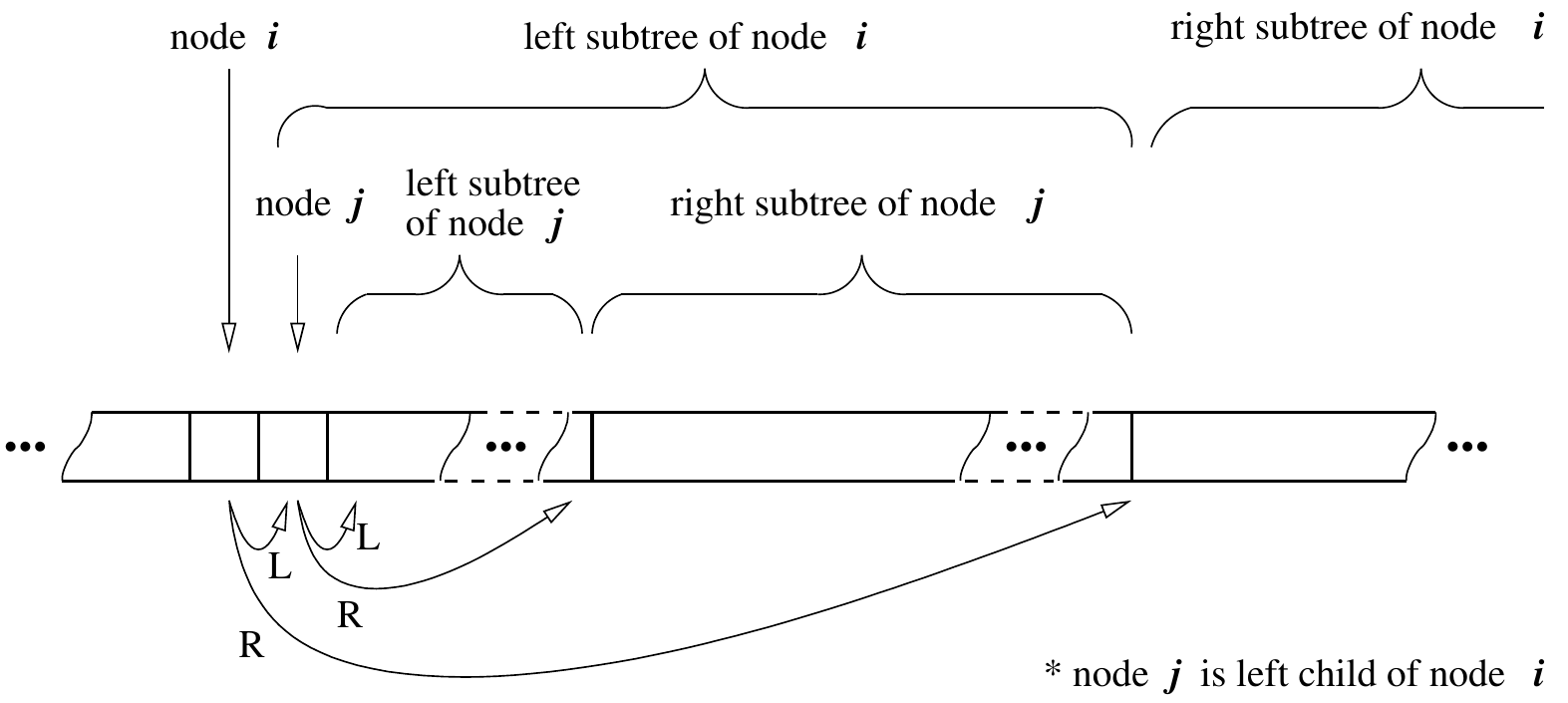}
\caption{Storage of an implicit IOT as a vector.
  The whole left sub-tree is positioned immediately after the parent
  node.  Right sub-tree (with missing root, because it is an implicit
  node) follows immediately after the left one.  This is repeated
  recursively.}
\label {implicit-iot-in-a-vector} 
\vspace{-10pt}
\end{figure}

There is no need for special memory allocations and storage of
pointers to child nodes.  The size of the (real part of) left sub-tree
under each particular node can be found (for maximal gray value
precision of 1) as $2^{d-i}-1$, where $i$ is depth of the node in a
tree.  Positioning the whole left sub-tree immediately after the
parent node and applying this rule recursively, all nodes of any
implicit uniform interval-occurrences tree can be stored in a vector
and then accessed by indexing operations
(Fig. \ref{implicit-iot-in-a-vector}).  Such data organization also
improves memory access pattern and cache utilization because data are
accessed in sequential order with strictly increasing addresses (array
indexes).

One can see that the use of implicit IOT reduces memory requirements
in approximately two times, down to the same value as a plain
single-layer histogram would require.  This happens without any
performance penalties. Instead a few operations during value
additions/removals are even saved.
\footnote{ The top-most node always equals to the
  total number of values saved in the IOT.  Since this is a known and
  for running window a constant value, storage of the top-most node can be
  avoided too.
  This saves nearly no space, but reduces IOT hight by 1 in each
  value insertion or removal as well as in each statistic request and
  window IOT update (what is much more interesting).}

\subsection{Insertion and removal of pixel values in an IOT}

Insertion (addition) of a value to an implicit uniform IOT can be done as follows
({\em add\_value} procedure): 
\begin{enumerate*}
\item Make top-most node the current one. Increment by one occurrences stored here.  
\item Make left child node the current one.
\item \label{add-value-algo-loop} If the value to be inserted is lower
  than the upper bound of the current node, then increment by one
  occurrences stored here and descend into the left
  sub-tree, i.e. make left child node of the current node the new current one.
  \\Otherwise: consider the {\em implicit} right node on the same level,
    which corresponds to the current (left) one
    (Fig.~\ref{explicit-iot-example}); incrementation of occurrences
    counter is not necessary since the node is an implicit one;
    just descend into the sub-tree growing from the implicit node,
    i.e. make left child node of the implicit node the current one.
\item If value interval size of the current node is greater than the
  predefined precision of the particular IOT, e.g. 1, then go to step
  \ref{add-value-algo-loop}.  Otherwise: finish.
\end{enumerate*}

Removal of a value is identical to the insertion with the only difference
that the respective occurrences are decremented instead of being
incremented.  Both insertion and removal are constant time operations
and require $d$ comparisons and in average $d/2$ additions
per operation.

\subsection{Adaptive interval-occurrences tree}
\label{adaptive-iot-section}

The number of additions and comparisons the algorithm makes directly
depends on the tree height, which is 9 for a uniform IOT and 8-bit
data and 17 for the 16-bit data.  This height is constant for all
values counted by a uniform interval-occurrences tree.  It is,
however, possible to build an interval-occurrences tree which has
lower height for values which occur more frequently and allow a higher
tree for seldom values, minimizing average height in this way.  This
is the same idea as used in entropy coding.

Let $T$ be a topology of interval-occurrences tree.  Then the
criterion for selection of the optimal $T$ can be formulated as
\begin{equation*}
\min_T{\sum_{op \in OP}{p(op) h(op)}},
\end{equation*}
where $op$ is an operation for insertion/removal of a specific value
or a calculation of a specific order statistic, $p(op)$ is relative
frequency of occurrence of operation $op$, $h(op)$ is height of the
IOT, i.e. number of nodes, visited to accomplish the request and the
summation is performed over all operations $OP$ performed by filtering
of an image.  Note that $h(op)$ is defined by the tree topology $T$
and $T$ is the same for the window IOT as well as for the column IOTs.
Thus, it has to be optimized for the value insertion/removal as well
as for order statistic calculation at the same time.

Variable height can be implemented by allowing child sub-intervals to
divide their parent's interval into non-equal parts.  We give it
without proof that the optimization criterion is met if such sizes of
sub-intervals are chosen that both children are visited equally frequent
in course of algorithm execution. 

The possibility of variable value intervals and variable height are
properties which distinguish the IOT data structure from the
conventional multi-tier histograms.  That is why we
give it a different name in general.  The version utilizing the
variable height is particularly called {\em adaptive
  interval-occurrences tree}.

\begin{figure} 
\centering \includegraphics[width=0.95\linewidth]{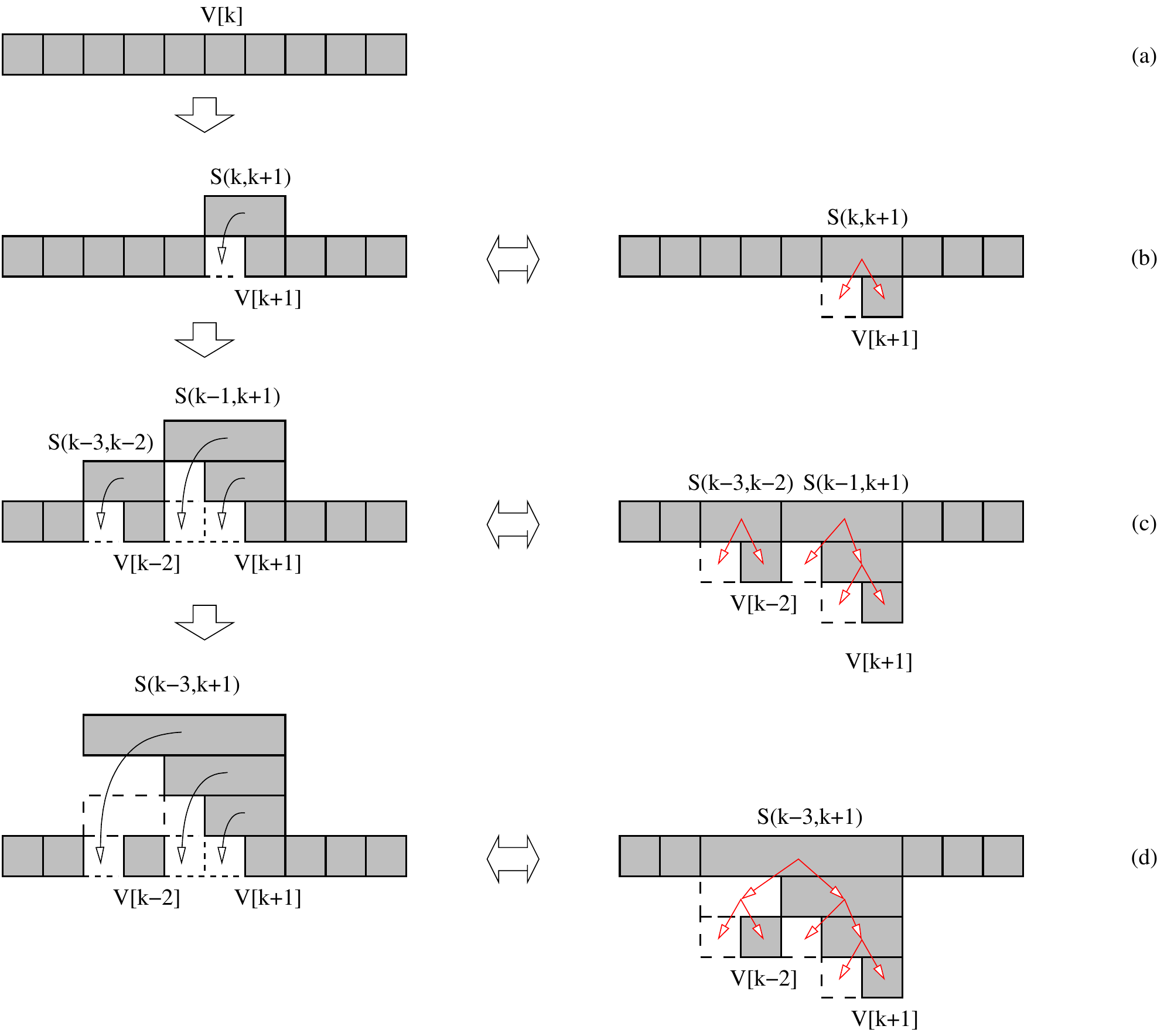}
\caption{An example of hierarchical vector representation using
  implicit elements (shown dashed): (a) original data vector; (b)-(d)
  successive stages of conversion to a hierarchical representation.
  Obviously, storage space requirements remain constant and equal to
  the size of the original (plain) data vector independent of selection
  of elements which are grouped at next representation stage.
  Implicit elements can be always (recursively) recomputed from the higher
  element and its corresponding neighbor.}
\label {adaptive-iot-space-requiremens}
\vspace{-10pt}
\end{figure}

Storage requirements of an adaptive implicit IOT remain the same as
for the uniform one or the plain histogram.  This follows from the
following theorem: 
any data vector can be stored in a hierarchical structure
requiring the same space
as the original vector independently of the hierarchy's
topology.  \begin{IEEEproof} Consider a data vector $V$ of finite size $N$
  (Fig.~\ref{adaptive-iot-space-requiremens}.a).  An arbitrary element
  $V[k]$ can be made implicit and expressed via its sum with the
  following or the preceding element of $V$, e.g.  $V[k]= S(k,k+1) -
  V[k+1]$, where $S(k,k+1)=\sum_{i=k}^{k+1}V[i]$.  Thus, $S(k,k+1)$
  can be stored instead of $V[k]$, see
  Fig.~\ref{adaptive-iot-space-requiremens}.b.  Obviously, such
  representation still requires the same storage $N$. This can be
  repeated recursively for any $S(.,.)$ or another element of $V$,
  excluding those which are already used in some $S(.,.)$ (e.g. as in
  Fig.~\ref{adaptive-iot-space-requiremens}.c,d), until $S(1,N)$ is
  created.  Then, one of many possible hierarchical representations is
  build.  Storage requirements for any configuration and at any step
  are constantly $N$.
\end{IEEEproof}

Topology of the adaptive IOT, i.e. lower and upper bounds of all
subintervals, has to be stored only once for all IOTs used in the
particular filter.  Thus the added space requirements are negligibly
small.
For the optimal IOT partitioning the frequencies have to be known at
which particular nodes, i.e. the associated value intervals, are
accessed.  Value insertion and removal operations access intervals
which correspond to values being inserted or removed.  Thus, a
conventional {\em global} histogram provides the necessary frequency
information.  On the other hand, the nodes accessed by the {\em
  get\_order\_statistic} procedure correspond to values of resulting
median.
Although they are generally not a priori known, a smoothed histogram
of the median of a typical image is sufficient for this purpose.  
A running {\em mean} (its global histogram) can be also used as a
first approximation, which is an $O(1)$ operation too.  In video
processing, the median distribution of the previous frame can be used.

Exactly one value insertion and one value removal is performed per one
pixel of result image.
Thus, we mix the source image global histogram and the global
histogram of median estimation at 1:1 ratio.  In case 
this mix vanishes for some value interval,
it is just divided to equal parts, same as for the uniform IOT.  Thus,
without a priori knowledge about source data and filter result an
adaptive IOT smoothly transforms into a uniform one.

\subsection{Variable precision}

If the IOT-based median calculation does not descend down to leafs,
which correspond to highest precision, but stops sooner, 
the result of median calculation is still meaningful.  
It just does not exhibit the maximal possible precision.  
The maximal error bound is also known: it is the width of the 
value interval corresponding to the last visited node.  
This allows to control the precision dynamically and
independently for each image point.  
Important effect of the reduced precision on the algorithm performance is
the avoidance of expensive non-elementary on-demand updates which occur for
deeper nodes in the window IOT.

\section{Experimental evaluation}

The described algorithm for calculation of running median is
implemented in C/C++.  The source code is available under GPL from the
algorithm web site 
(\url{http://helios.dynalias.net/~alex/median}).  In course of
experimental evaluation, the uniform IOT and the adaptive IOT versions
of algorithm were compared to OpenCV implementation of~\cite{Perreault2007}
(\url{http://opencv.willowgarage.com/}).  Special code snippets were
included in implementation of both algorithms per conditional
compilation for purpose of counting of additions and comparisons
inherent to each algorithm.  This way the data dependent number of
operations performed in reality can be evaluated. Results of this
evaluation for 8-bit data along with measurements of execution time
(without this additional counting code) are given in
Table~\ref{8bitPerformanceTab}.\footnote{ Experimentally measured number of
  additions and comparisons for the maintenance of column statistics
  are slightly different from theoretical values
  (being for 8-bit data 4 additions for~\cite{Perreault2007} 
  and 8 additions and 16 comparisons for
  the uniform IOT version of the proposed algorithm).  
  This is due to 
  border effects.
  }  Evaluation results for 16-bit data
are given in Table~\ref{16bitPerformanceTab}, although a comparison to
Perreault and H\'{e}bert algorithm was not possible because of absent
implementation of this algorithm for 16-bit data.
Table~\ref{16bitPerformanceTab} also includes results for reduced
precision calculation on 16-bit data.  Measurements of execution time
are represented in number of CPU clock ticks utilized per output
pixel.  Evaluation is performed on Intel Core i7 M620 CPU (single core
is used), code was compiled with gcc-4.4 without
as well as with 
SIMD optimization.

\begin{figure} 
\centering \includegraphics[width=0.95\linewidth]{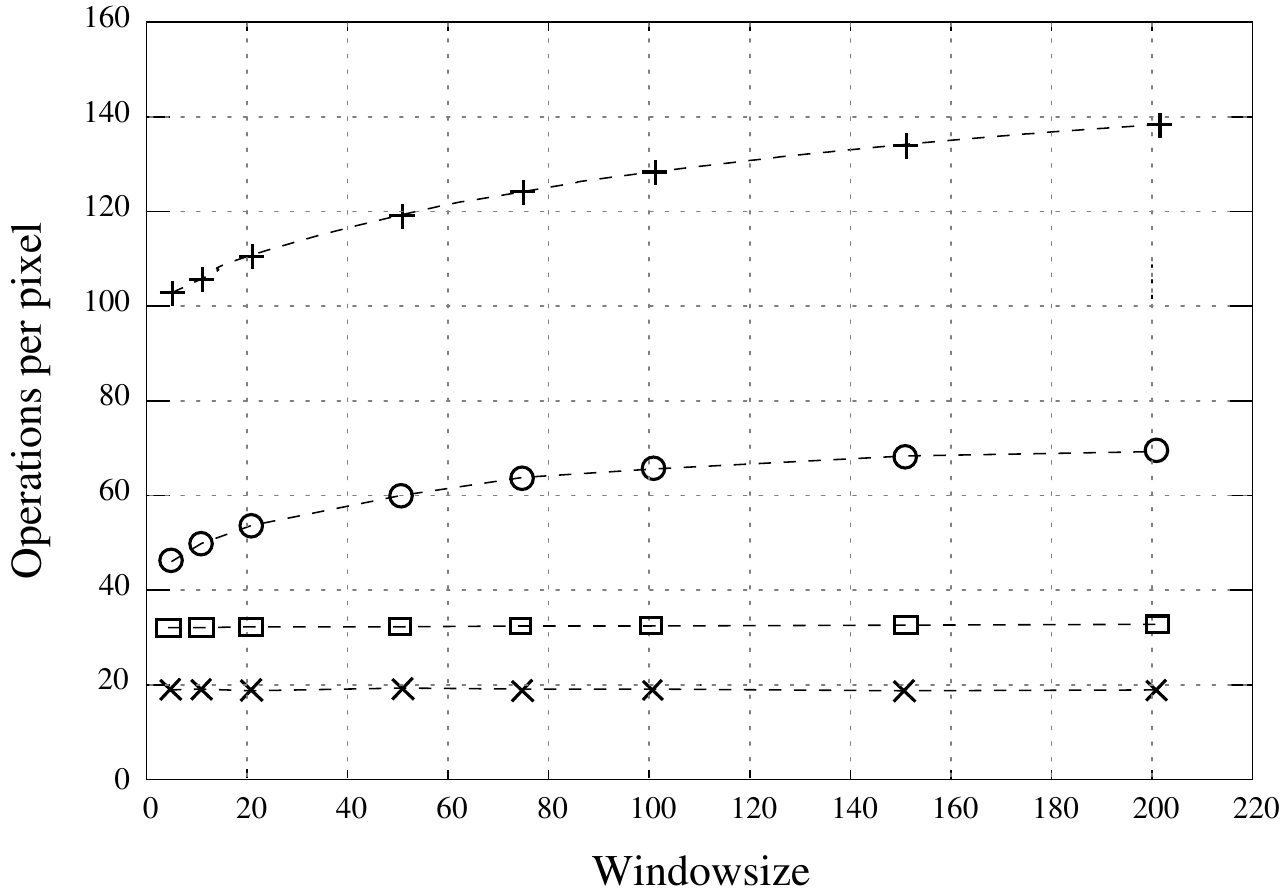}
\caption[Operations vs. filter size]{Average number of operations per output pixels made by the
  Perreault and H\'{e}bert~\cite{Perreault2007} and the developed algorithm as
  function of filter size. 16 MPix ``airfield'' image
  (Fig. \ref{test-images}.a) is used. $+$ and $\times$ - additions and
  comparisons by Perreault, {\small $\bigcirc$} and $\square$ -
  additions and comparisons by the uniform IOT version of the
  developed algorithm.}
\label {performance-for-var-windowsize}
\vspace{-10pt}
\end{figure}

\begin{figure*} 
\centering
\begin{tabular}{cccc}
\includegraphics[height=0.24\linewidth]{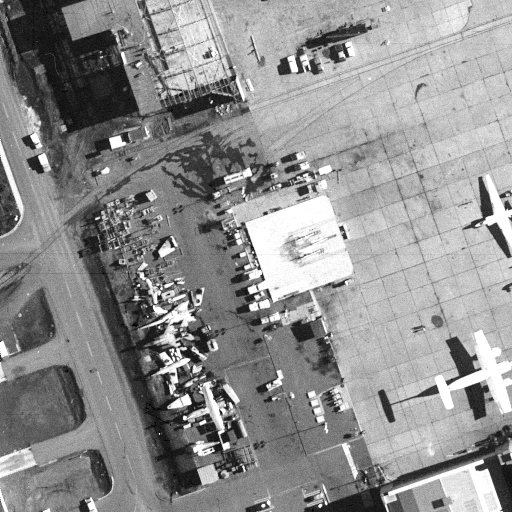} &
\includegraphics[height=0.24\linewidth]{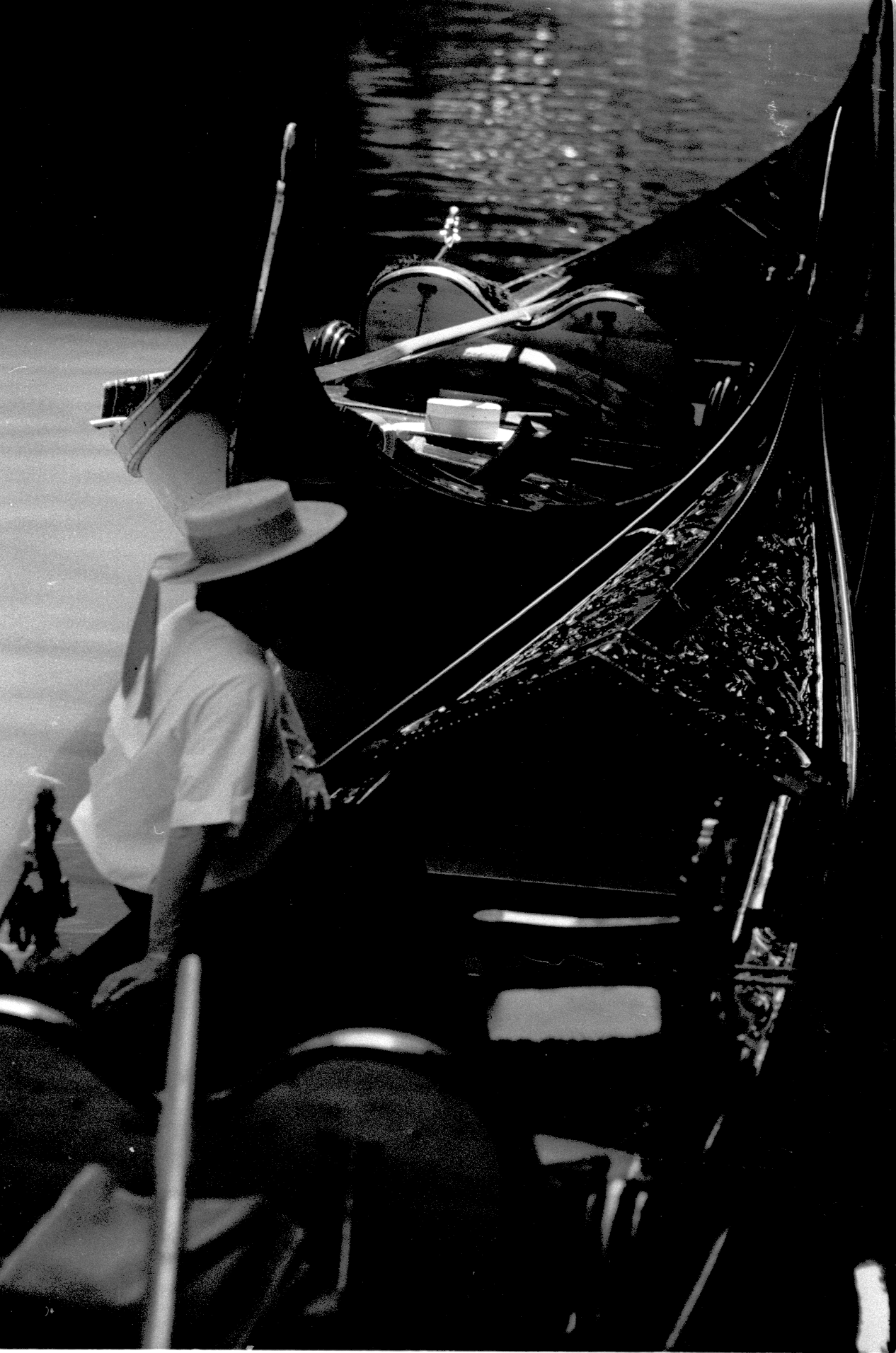} &
\includegraphics[height=0.24\linewidth]{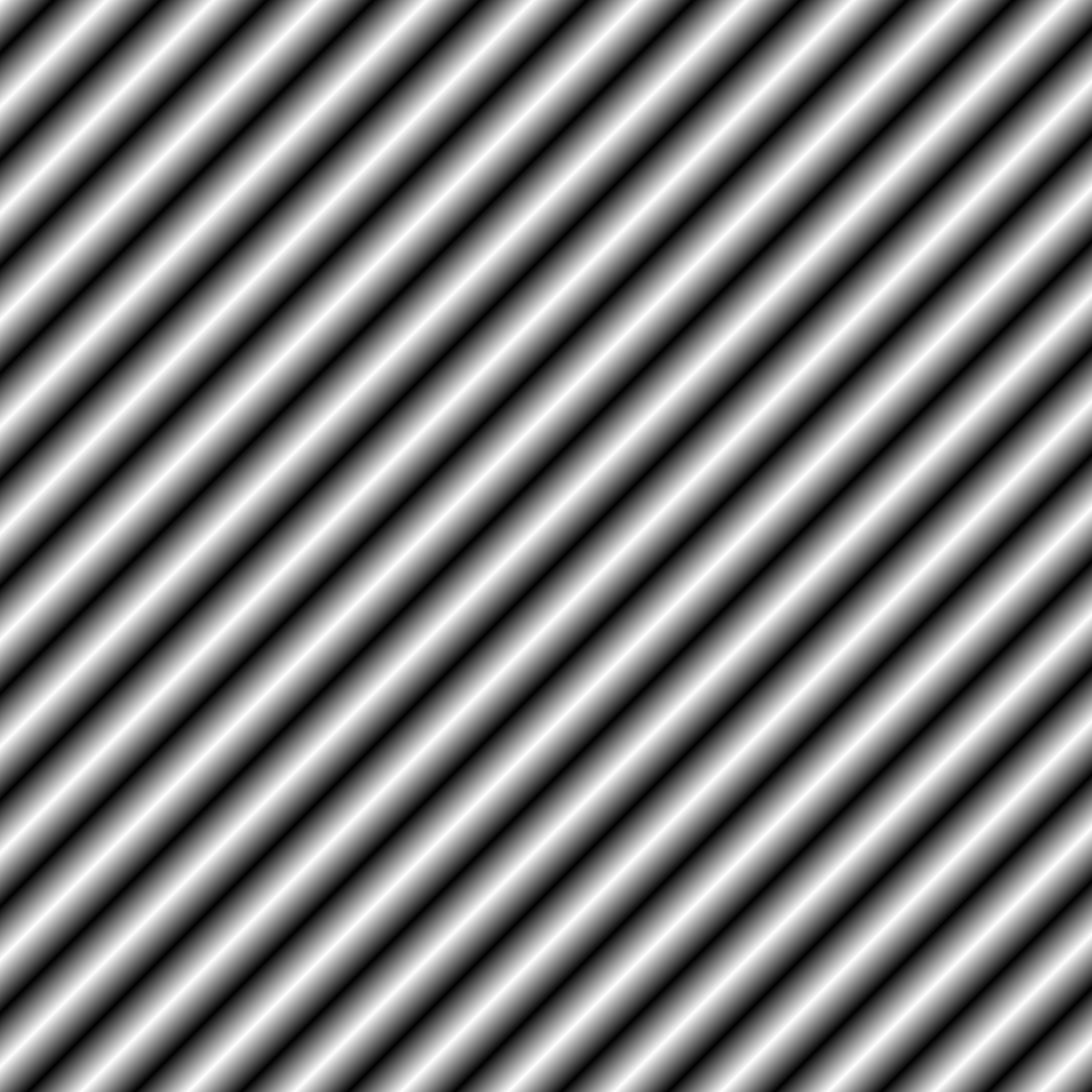} &
\includegraphics[height=0.24\linewidth]{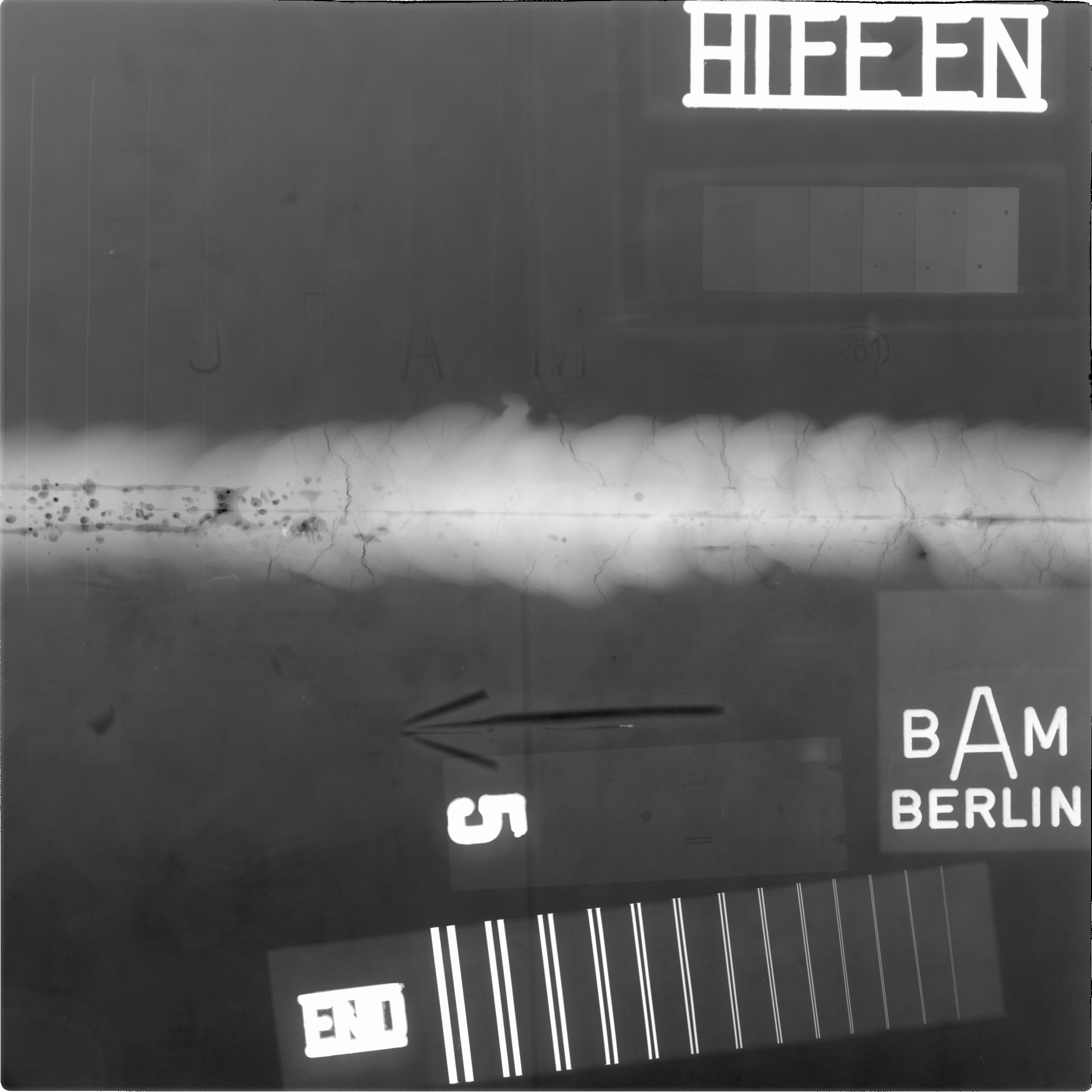} \\
a) & b) & c) & d) \\
\end{tabular}
\caption[Test images]{Test images used for evaluation: 
  a) classic ``airfield'' image up-scaled to 4096 x 4096 with some 
  uncorrelated uniformly distributed noise added to create 
  high-frequency components (gray values fill whole 0..255 range 
  and the histogram is fairly uniform); 
  b) a low-key photograph (an 8-bit image with compact histogram); 
  c) artificial worst case image similar to one used in~\cite{Perreault2007}; 
  d) X-ray inspection of a welding (a 16-bit image);
  }
\label{test-images}
\vspace{-10pt}
\end{figure*}

The above comparison was done on various source images
(Fig.~\ref{test-images}) for one fixed size of the filter window.
Fig.~\ref{performance-for-var-windowsize} shows algorithms' performance
for variable window size on example of one arbitrary image
(``airfield''), demonstrating approximately constant computational complexity
of both algorithms in the practically relevant window size range.

We purposely do not show here any filter results because they are the
same irrespective of the applied algorithm.

We have also tested the {\em Quickselect}-based median filter on
16-bit data,
as the most used one for non-8-bit data.  It was slower than the
proposed algorithm for window sizes larger than 4x4 pixels.  Particularly,
for window sizes 11x11 and 51x51 it was slower in approximately 3.9 and 43 
times respectively.

\begin{table*}
\ifCLASSOPTIONdraftcls
\renewcommand{\arraystretch}{0.8}
\else
\renewcommand{\arraystretch}{1.25} 
\fi
\caption[Performance for 8-bit images] {\parbox[t]{0.85\linewidth}
  {\rm Operations breakdown (per pixel) and performance of Perreault
    and H\'{e}bert algorithm~\cite{Perreault2007} and the uniform and
    adaptive versions of the developed algorithm for different source
    data.  Filter window size $n$ is fixed to 51 pixels, because
    algorithm' complexities are roughly independent of $n$.  Additions
    and comparisons are listed separately.  Runtime$^\ast$ is number
    of CPU clock ticks utilized per pixel, Intel Core i7 M620 CPU
    (single core is used), compiled with gcc-4.4 with {\tt
      -00~-fno-tree-vectorize}.  Runtime$^{\ast\ast}$ is measured
    under the same conditions but compiling with {\tt
      -02~-ftree-vectorize}. ``Winner'' is shown in bold font.}}
\label{8bitPerformanceTab}
\centering 
\footnotesize
\begin{center} \begin{tabular}{|c|*{5}{|r|r|}} %
\hline
Algorithm & 
  \multicolumn{2}{c||}{Maintenance of columns} &
  \multicolumn{2}{c||}{Update of window} & 
  \multicolumn{2}{c||}{Median extraction} & 
  \multicolumn{2}{c||}{Overall per pixel} &
  \parbox[t]{0.05\linewidth}{Run\-time$^\ast$} &
  \parbox[t]{0.05\linewidth}{Run\-time$^{\ast\ast}$} \\ 
\cline{2-9}
 & ~~~~add~~~ & ~~~cmp~~~ & ~~add~~ & ~~cmp~~ & ~~add~~ & ~~cmp~~ & ~~add~~ & ~~cmp~~ & & \\
\hline

\multicolumn{11}{l}{} \\
\multicolumn{11}{l}{Random uncorrelated normally distributed values,
$\mu=128$, $3\sigma=128$, 4096 x 4096 image}\\ 
\hline
Uniform~ IOT            &  8.1 & 16.2 & 31.2 & 8.0 &  8.0 &  8.0 &  47.3 & 32.2 & \bf{623} &     233  \\
Adaptive IOT            &  8.4 & 36.0 & 16.1 & 4.0 &  4.6 &  7.9 &  29.1 & 47.9 &     717  &     333  \\ %
Perreault \& H\'{e}bert &  4.4 &  0.0 & 86.6 & 1.0 & 21.2 & 19.2 & 112.2 & 20.2 &    1435  & \bf{159} \\
\hline

\multicolumn{11}{l}{} \\
\multicolumn{11}{l}{Real image with equalized histogram
(classic ``airfield'' target up-scaled to 4096 x 4096)} \\ 
\hline
Uniform~ IOT             &  8.1 & 16.2 & 44.0 & 8.0 &  8.0 &  8.0 &  60.1 & 32.2 & \bf{744} &     280  \\
Adaptive IOT             &  8.1 & 33.0 & 44.7 & 8.2 &  8.8 & 16.4 &  61.6 & 57.6 &     1000 &     406  \\ %
Perreault \& H\'{e}bert  &  4.4 &  0.0 & 94.3 & 1.0 & 20.0 & 18.0 & 118.7 & 19.0 &    1543  & \bf{170} \\ 
\hline

\multicolumn{11}{l}{} \\
\multicolumn{11}{l}{Real image with compact histogram 
  (``Venice at night'', 2532 x 3824)} \\ 
\hline
Uniform~ IOT            & 8.2 & 16.3 & 29.6 & 8.0 & 8.0 & 8.0 & 45.7 & 32.3 &     548  &     144  \\
Adaptive IOT            & 4.0 & 14.2 & 14.7 & 3.2 & 4.2 & 6.4 & 22.8 & 23.9 & \bf{368} & \bf{126} \\ %
Perreault \& H\'{e}bert & 4.4 &  0.0 & 78.9 & 1.0 & 7.3 & 5.3 & 90.5 &  6.3 &    1192  &     129  \\
\hline

\multicolumn{11}{l}{} \\
\multicolumn{11}{l}{Worst case peak: synthetic diagonal sine pattern with 100 pix period} \\ 
\hline
Uniform~ IOT            &  8.1 & 16.2 & 100.3 & 8.0 &  8.0 &  8.0 & 116.4 & 32.2 & \bf{1219} &     365  \\
Adaptive IOT            &  9.2 & 34.6 & 100.8 & 8.4 &  9.4 & 16.9 & 119.4 & 60.0 &     1508  &     490  \\ %
Perreault \& H\'{e}bert &  4.4 &  0.0 & 165.3 & 1.0 & 19.0 & 17.0 & 188.7 & 18.0 &     2495  & \bf{223} \\
\hline

\multicolumn{11}{l}{} \\
\multicolumn{11}{l}{Synthetic diagonal sine pattern with 25 pix period} \\ 
\hline
Uniform~ IOT            &  8.1 & 16.2 & 27.1 & 8.0 &  8.0 &  8.0 & 43.2 & 32.2 &     490  & \bf{114} \\
Adaptive IOT            &  7.8 & 28.5 &  9.9 & 3.0 &  3.9 &  6.0 & 21.7 & 37.5 & \bf{482} &     168  \\ %
Perreault \& H\'{e}bert &  4.4 &  0.0 & 77.2 & 1.0 & 12.9 & 10.9 & 94.5 & 11.9 &    1219  &     136  \\
\hline

\end{tabular} \end{center} 
\end{table*}

\begin{table*} 
\ifCLASSOPTIONdraftcls
\renewcommand{\arraystretch}{0.8}
\else
\renewcommand{\arraystretch}{1.25} 
\fi
\caption[Performance for 16-bit images] {\parbox[t]{0.85\linewidth}
  {\rm Operations breakdown and performance of uniform and adaptive
   versions of the developed algorithm on 16-bit data.  Performance
   of uniform IOT for precision reduced down to 16 gray values
   included (``UIOT, err16'').  Comparison to Perreault \& H\'{e}bert
   is not possible because of missing implementation.  Filter size is
   51 and other conditions are the same as for
   Tab.~\ref{8bitPerformanceTab}.  Additionally, the performance in
   millions of pixels per second is given (single core of Intel Core
   i7 M620 CPU @ 2.67 GHz)} }
\label{16bitPerformanceTab} 
\centering
\footnotesize
\begin{center} \begin{tabular}{|c|*{5}{|r|r|}} %
\hline
Algorithm & 
  \multicolumn{2}{c||}{Maintenance of columns} &
  \multicolumn{2}{c||}{Update of window} & 
  \multicolumn{2}{c||}{Median extraction} & 
  \multicolumn{2}{c||}{Overall per pixel} &
  \parbox[t]{0.06\linewidth}{Run\-time$^{\ast\ast}$} & 
  \parbox[t]{0.05\linewidth}{Mp/s} \\ 
\cline{2-9}
 & ~~~~add~~~ & ~~~cmp~~~ & ~~add~~ & ~~cmp~~ & ~~add~~ & ~~cmp~~ & ~~add~~ & ~~cmp~~ & & \\
\hline

\multicolumn{11}{l}{} \\
\multicolumn{11}{l}{Random uncorrelated normally distributed gray values, $\mu=2^{15}$, $3\sigma=2^{15}$} \\ 
\hline
Uniform~ IOT & 16.2 & 32.4 & 208.8 & 16.0 & 16.0 & 16.0 & 241.0 &  64.4 & 2025 & 1.32 \\
UIOT, err16  & 16.2 & 32.4 &  80.4 & 12.0 & 12.0 & 12.0 & 108.6 &  56.4 & 1510 & 1.77 \\
Adaptive IOT & 16.9 & 68.2 & 194.8 & 11.7 & 12.3 & 23.4 & 224.0 & 103.3 & 3180 & 0.84 \\ %
\hline

\multicolumn{11}{l}{} \\
\multicolumn{11}{l}{Real 16-bit image (industrial X-ray inspection of weldings)} \\ %
\hline
Uniform~ IOT & 16.8 & 33.6 & 137.5 & 16.0 & 16.0 & 16.0 & 170.3 & 65.6 &  931 & 2.87 \\
UIOT, err16  & 16.8 & 33.6 &  52.2 & 12.0 & 12.0 & 12.0 &  81.0 & 57.6 &  638 & 4.18 \\
Adaptive IOT & 13.5 & 54.6 & 129.4 & 12.9 & 13.5 & 25.8 & 156.5 & 93.3 & 1010 & 2.64 \\ %
\hline

\multicolumn{11}{l}{} \\
\multicolumn{11}{l}{Worst case peak: synthetic diagonal sine pattern with 100 pix period} \\ 
\hline
Uniform~ IOT & 16.2 & 32.4 & 477.5 & 16.0 & 16.0 & 16.0 & 509.7 &  64.4 & 4600 & 0.58 \\
UIOT, err16  & 16.2 & 32.4 & 278.1 & 12.0 & 12.0 & 12.0 & 306.3 &  56.4 & 2935 & 0.91 \\
Adaptive IOT & 16.2 & 65.6 & 474.0 & 16.2 & 16.8 & 32.3 & 506.9 & 114.1 & 5040 & 0.53 \\ %
\hline

\end{tabular} \end{center} 
\end{table*}

\section{Analysis}

\subsection{Vectorized calculations
}

SIMD (single instruction, multiple data) is paralleling scheme
implemented in hardware of many modern main stream 
CPUs.  Using SIMD extensions several identical operations on vectors
of data, e.g. several additions, subtractions or logical operations,
can be performed in parallel in one execution step.
Weiss~\cite{Weiss2006} and Perreault and
H\'{e}bert~\cite{Perreault2007} propose to utilize SIMD instructions
sets found on modern CPUs and accelerate running median calculation in
this way.  Their algorithms, based on plain histograms, benefit naturally of
these possibilities.

The last two columns of Table \ref{8bitPerformanceTab} show how the
execution time of Perreault and H\'{e}bert
algorithm~\cite{Perreault2007} can be improved thanks to SIMD and
other hardware-specific optimizations.  Although nearly twice as many
additions have to be made by this algorithm, they can be executed in
parallel, drastically reducing overall execution time.  Whereas the
developed algorithm requires in fact less operations as algorithm
in~\cite{Perreault2007}, it cannot benefit from this hardware
parallelism to the same great
extend because of significantly higher number of branches
(executed comparisons) in it.  
Nevertheless, the developed algorithm makes less
operations per pixel, offering therefore higher efficiency,
and will be faster on CPUs without high internal parallelism.

\subsection{High precision data}

The developed algorithm needs no modifications to be used on 16-bit images.
A uniform IOT for 16-bit data has height of 16.  It is twice as high
as for 8-bit data, thus the 
first expectation for increase in number of operations is also a
factor of two.  However, as Table~\ref{16bitPerformanceTab} shows,
significantly more non-elementary updates of window IOT are
required in practice.  Despite the algorithm remains roughly $O(1)$,
the increase in number of operations (for these data) is approximately a factor of 5.  
It is still an excellent result
since, for example, in Weiss's algorithm~\cite{Weiss2006} the
complexity would be squared.

We cannot perform experimental comparison to Perreault and H\'{e}bert
algorithm for 16-bit data, because the corresponding implementation
is not available.  Instead of two-tier histogram they propose to use a
three-tier or four-tier one for 16-bit data.  We expect, however, that
our approach will be superior because of fully optimized hierarchical
data organization.

\subsection{Reduced precision output}

Table \ref{16bitPerformanceTab} shows that performance is increased
significantly if precision requirements are slightly relaxed.  For this
evaluation on 65536-gradation gray value data we allow an error of
median estimation of $\simeq 0.02\%$ (i.e. 16 gray values).  The error
limit can be set for each individual pixel arbitrarily and
independently, but for simplicity of evaluation we just set the same
value for all pixels.  For allowed error of 16 gray values the
algorithm does not evaluate deepest 4 levels in the window IOT.
This way many of non-elementary updates of nodes in window IOT are
avoided (see ``update of window'' column of
Table~\ref{16bitPerformanceTab}).

\subsection{Data dependency and performance of adaptive IOT}

Adaptive version of the developed algorithm requires one more comparison 
per node visit. 
It is because the tree has variable height and it must be always
tested whether the node is a leaf.  This is not necessary for uniform
IOT where tree height is fixed and it is possible to determine in
advance through how many layers the algorithm has to descend in order
to reach leaf nodes.  Thus the possibility of adaptation comes at
added computational complexity costs (see operations breakdown in
Tables~\ref{8bitPerformanceTab} and \ref{16bitPerformanceTab}).

In order to be more efficient than the uniform version, the adaptive
IOT must offer operation savings which compensate for these additional
comparisons.  The best case for this arises if gray value distribution
of source image is highly correlated to gray value distribution of
filtered image and if they show distinctive compact areas where most
of gray values are located.  If this is not the case,
then the adaptive IOT version performs worse than its simpler uniform
one.  Thus, in our experiments the adaptive IOT version was able to
show better performance only for images with compact histograms like
the example in Fig.~\ref{test-images}.b.

\section{Conclusion and outlook}

A new approximately constant time algorithm for calculation of running
median and other order statistics is developed.  The algorithm is based
on a hierarchical data structure for storage of value occurrences in
specific value intervals.  The computational complexity of the
algorithm is
lowest among currently available algorithms.  Experimental comparison
to the single other constant time algorithm~\cite{Perreault2007} confirms this.  
The competitive algorithm can, however, compensate
its higher complexity by 
the extensive use of SIMD CPU extensions.
Therefore it was not possible to show any practically relevant
improvements of real-world execution time over it on 8-bit data if executed on a
modern SIMD-enabled CPU.

Nevertheless, the new algorithm will be interesting for higher
precision data and images with prevailing gray value ranges like HDR
and X-ray imaging, low/high-key images, surveillance video, etc. as
well as on platforms lacking hardware SIMD extensions, e.g. embedded
systems, mobile devices, etc.  Usual higher-level
parallelizations are naturally possible too.

The developed algorithm needs no modifications to be used on 16-bit images.
Straight-forward extension to 32-bit integer data becomes, however,
impractical because of high memory requirements.  Analogous, a general purpose
extension for floating point data is not possible.  
The proposed by Weiss ordinal transform~\cite{Weiss2006} can be used
to address both problems, but will probably hamper performance.
The usage of Chebyshev's inequality as in~\cite{Tibshirani2009} but in
the IOT framework seams a better idea.  This needs further
investigations.

\section*{Acknowledgments}

This work was done during author's engagement at the Computer
Vision and Remote Sensing Group of 
Technische Universität 
Berlin.  He would 
like to thank
Olaf Hellwich, 
Peter Le\v{s}kovsk\'{y} 
and Ronny H\"{a}nsch 
for their valuable suggestions and feedback.

\bibliography{median}{}
\bibliographystyle{plain}

\begin{IEEEbiography}
[{\includegraphics[width=1in,height=1.25in,clip,keepaspectratio]{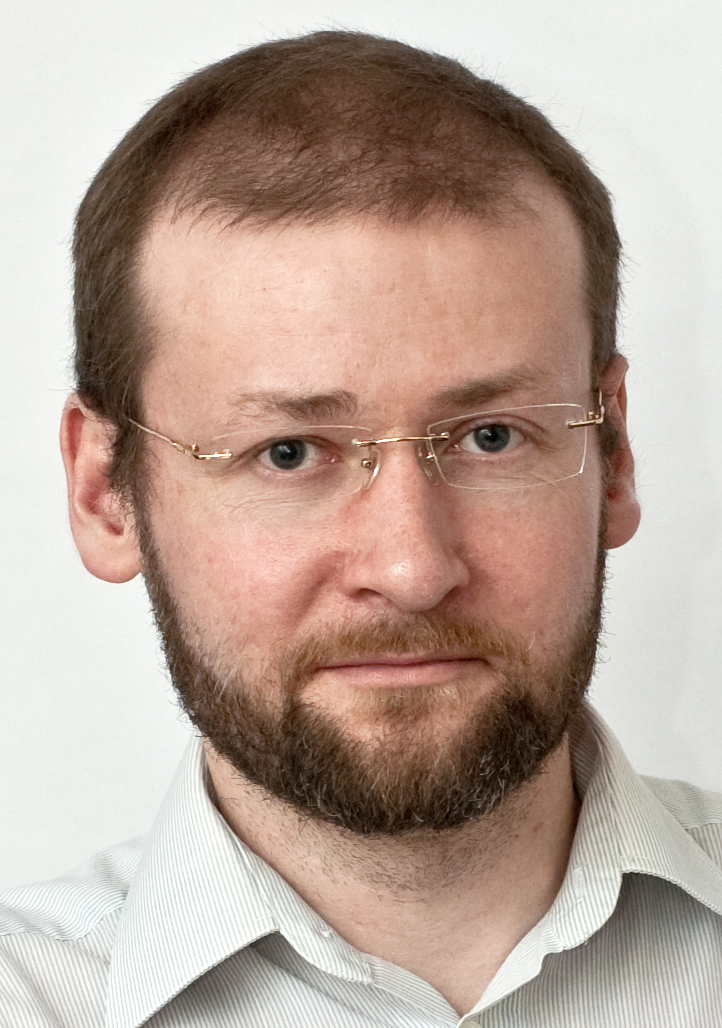}}]
{Alexander Alekseychuk} has got his MS degree in electrical
engineering from Lviv's Institute of Technology (Lvivska Polytechnika)
in 1994.  Till 2000 he was with the Institute of Physics and Mechanics
of the Ukrainian National Academy of Science.  In 2000 he attended the
German Federal Institute for Materials Research and Testing (BAM)
where he dealt with algorithm and software development for image
processing in application to digital industrial radiology.  He has got
his PhD degree for work in pattern recognition from the Technische
Universität Dresden in 2006.  Since 2010 he is with the Computer
Vision and Remote Sensing Group at the Technische Universität Berlin.
His former and current scientific interests are in the field of
efficient algorithms, object detection on busy backgrounds and under
low signal-to-noise ratios, texture-based segmentation and
content-based image retrieval.

\end{IEEEbiography}

\end{document}